\documentclass{article}\usepackage[]{graphicx}\usepackage[]{color}
\makeatletter
\def\maxwidth{ %
  \ifdim\Gin@nat@width>\linewidth
    \linewidth
  \else
    \Gin@nat@width
  \fi
}
\makeatother

\definecolor{fgcolor}{rgb}{0.345, 0.345, 0.345}

\usepackage{framed}
\makeatletter
 {\par\unskip\endMakeFramed%
 \at@end@of@kframe}
\makeatother

\definecolor{shadecolor}{rgb}{.97, .97, .97}
\definecolor{messagecolor}{rgb}{0, 0, 0}
\definecolor{warningcolor}{rgb}{1, 0, 1}
\definecolor{errorcolor}{rgb}{1, 0, 0}
\newenvironment{knitrout}{}{} 

\usepackage{alltt}

\usepackage[utf8]{inputenc}
\usepackage{hyperref}

\title{Measuring the local GitHub developer community}
\author{JJ Merelo$^{1,2,3}$\thanks{Corresponding author. He can be
    reached at {\tt jmerelo@ugr.es} or at the
    \href{https://github.com/geneura-papers/github-ranking/issues}{issues
      section} of the repo for this paper} \and Nuria Rico$^{1,2}$
  \and Israel Blancas$^1$ \and M. G. Arenas$^{1,3}$ \and
  Fernando Tricas $^4$ \and José Antonio Vacas\\
  $^1$University of Granada, Spain\\
  $^2$Free Software Office\\
  $^3$Geneura Team\\
  $^4$University of Zaragoza, Spain
}
\IfFileExists{upquote.sty}{\usepackage{upquote}}{}
\begin{document}
\date{}

\maketitle

\begin{abstract}
  Creating rankings might seem like a vain exercise in belly-button
  gazing, even more so for people so unlike that kind of things as
  programmers. However, in this paper we will try to prove how
  creating city (or province) based rankings in Spain has led to   
  all kind of interesting effects, including increased productivity and
  community building. We describe the methodology we have used to
  search for programmers residing in a particular province  focusing
  on those where most population is concentrated and apply different
  measures to show how these communities differ in structure, number
  and productivity. 
\end{abstract}

\section{Introduction}

One of the keys to create a community is to actually identify who is
part of it and how they participate. As part of the effort by the Free Software Office of the
University of Granada, we have tried, through the years, to know who
is involved in the creation of open source projects. However, the only
way of finding out who was is to make them come to any of our events
or contact us through any means.

So the initial intention for creating a ranking of FLOSS (Free/Libre/Open/So\-ur\-ce Software)
was to know who is out there and the kind of things they are doing, be
them part of the academic world or outside it, in business; creating a census would allow us
to discover new FLOSS developers in our own city and even to collaborate with them.

So we used initially the GitHub top-1000 generation script by Paul
Miller \href{https://github.com/paulmillr/top-github-users} to achieve
that, making small modifications to the source and creating our own
version, which was eventually 
\href{https://github.com/JJ/github-city-rankings}{moved to a new repository}. But this had
several effects. First, as soon as the ranking was published some
people contacted us and the first GitHub meet-up in Granada took
place. More modifications and changes were added and new data was
obtained. As part of the code tests, more cities were tested and we
ended up with lots of data. And data begs for analysis, which we
eventually started to do. And, along the way, we built a community of users that previously had not known each other. We discovered that
the only fact that a census exists does not imply that there is a
community, but it definitely helps. 
We have had some experience with this kind of reactions in the past. In~\cite{blogtalk} we did some studies on social network analysis and other measures of the Spanish-speaking blogosphere. Then, the reactions were two fold: on one side, people showed big interest in the index in order to be listed there; some blog providers provided also data. On the other side, people that were expecting to appear in better positions were a bit angry about it\footnote{You can see some of the discussions and links -unfortunately most of them do not work- at: \url{http://www.blogalia.com/historias/7744} (in Spanish)}. Anyway we feel that self-consciousness is always a good thing and this work can serve as a driving force for more code sharing and increase relations among developers.

In this paper, using the tool that we have created for searching for
the users in particular cities or provinces, we will show how the
GitHub activity in these cities or provinces compare with each other
and what kind of characteristics they have, including basic
metrics. We will also delve into the effect of publishing the ranking
itself, which has surprisingly increased the productivity of all
communities measured. Finally, we will try to draw some conclusions on
how measuring activity affects that activity and what are the general
characteristics of open source developers in the provinces measured,
which are the top 20 in population in Spain.

Coming up next we will review different papers that deal with creating
lists and rankings of contributions and trying to measure or explain
the dynamics of communities. In section \ref{sec:exp} we will show the
methodology for obtaining the users in a particular province in Spain;
next we will analyze data obtained and show how different provinces
stack up in terms of contributions and finally we will draw some
conclusions. 

\section{State of the art}

Geographically based community metrics have had some attention in the last years~\cite{BarahonaRoblesAndradasGhosh08,TakhteyevHilts10,EngelhardtFreytagSchulz13} but they do not seem to have arisen a
lot of interest in the FLOSS metrics~\cite{herraiz2009flossmetrics} community. Most efforts seem to
be focused in creating tools for actually measuring repositories for
activity; for instance, Laura Arjona describes the {\em Debian
  Contributors} tool in \cite{arjona:debian} whose results are dumped
to a website that includes information on the projects that every user
has contributed and some other data such as how users are identified
in the databases. 
However, geography seems to be relevant in human interactions even when we are in Internet, where one could expect this factor to be less important. See, for example, `Visualizing Friendships'\footnote{\url{https://www.facebook.com/notes/facebook-engineering/visualizing-friendships/469716398919}} where the authors studied interactions inside the Facebook social network, or~\cite{RattiSobolevskyCalabreseAndrisReadesMartinoClaxtonStrogatz10} where the subject of analysis are phone calls. We can see that even when there is technology-mediated communication, the geography seems to be an important driver.

It is interesting to note that some models \cite{robles05} use the
concept of stigmergy, that is, interaction using the environment, to
model the dynamics of libre software projects; the mere existence of
these tools can be a catalyst of this interaction and the harbinger
of new software projects. In fact, this seems to be what has happened
in the community (or communities) under observation: the mere creation of
a document that mentions many different users acts as a substrate that
allows the creation and growth of the community through the stigmergy paradigm.

Next we will briefly explain the tool that was designed to search for
geographically based GitHub users.

\section{GitHub city rankings, the tool}
\label{sec:exp}
\begin{table}[htb]
    \centering

\begin{tabular}{l|r|r|r|r|r}
\hline
province & population & users & contributions & stars & followers\\
\hline
Alicante & 1852789 & 52 & 4941 & 255 & 276\\
\hline
Asturias & 1054408 & 59 & 8121 & 584 & 358\\
\hline
Baleares & 1121739 & 31 & 2037 & 361 & 212\\
\hline
Barcelona & 5435373 & 808 & 108576 & 35070 & 16836\\
\hline
Bilbao & 1138090 & 84 & 9071 & 1912 & 1475\\
\hline
Cádiz & 1247884 & 44 & 2497 & 604 & 401\\
\hline
Córdoba & 796680 & 65 & 3821 & 298 & 371\\
\hline
Coruña & 1130354 & 60 & 4551 & 1049 & 462\\
\hline
Gerona & 741017 & 29 & 2078 & 894 & 346\\
\hline
Granada & 919663 & 182 & 29610 & 1416 & 1243\\
\hline
Las Palmas & 1102750 & 56 & 3031 & 548 & 298\\
\hline
Madrid & 6376610 & 798 & 143739 & 37003 & 13375\\
\hline
Málaga & 1626168 & 86 & 7356 & 926 & 528\\
\hline
Murcia & 1463797 & 37 & 2893 & 752 & 272\\
\hline
Pontevedra & 948588 & 55 & 2939 & 1065 & 417\\
\hline
Sevilla & 1937412 & 115 & 11385 & 1255 & 1186\\
\hline
Tarragona & 792868 & 21 & 1353 & 160 & 121\\
\hline
Tenerife & 1017785 & 60 & 5816 & 577 & 591\\
\hline
Valencia & 2521771 & 215 & 20037 & 2718 & 1383\\
\hline
Zaragoza & 967354 & 86 & 13938 & 1466 & 1010\\
\hline
\end{tabular}

\caption{Raw aggregate measures for the 20 most populated provinces,
  including the population (taken from the National Statistics
  Institute), the number of users and their contributions, stars and
  followers. Please note that province names do not correspond to
  official names, having rather been chosen a bit arbitrarily from the
  search strings used.} 
\label{tab:data}
\end{table}
It is quite unlikely that if you are reading this you do not know what
is \href{http://github.com}{GitHub}. GitHub is a web-based git
repository that has a number of ``social'' features, including the
declaration of a profile and {\tt @} mentions in commit messages and
issues. The profile page includes information on the number of
followers, as well as the repositories and the number of contributions
every person has made during the last year. Besides this
easily-scrapeable information, GitHub has a REST API that can be
accessed from any language. 

Some other web-based \emph{repos} do have many of the characteristics, and,
besides, are based in free software themselves, like
Gitorious\footnote{\url{http://gitorious.org/}}, SourceForge\footnote{\url{http://sourceforge.net/}}, Google Code\footnote{\url{http://code.google.com/}}, just to name a few of them. However the number (and the activity) of users of these repositories is quite small compared to GitHub, which has become the tool of choice for
FLOSS developers. That is why GitHub was chosen, apart from the availability of tools to mine profile information: it provides an API that allows us to study things in an easier way. 
Notice also that all the projects in GitHub can not be considered FLOSS as we can see at~\cite{Williamson13}.
Nevertheless, people seems to be following the web culture where sharing and broadcasting are the usual ways of relation and they are not paying attention to licensing issues.

The \href{https://github.com/paulmillr/top-github-users}{tool used initially, by Paul
Miller} was
written in CoffeeScript and designed for creating a ranking of the top
1000 users with more than an (arbitrary) numbers of followers equal to
256. The tool used the GitHub REST API to make requests, and saved
them in a human-readable form in Markdown and also CSV and JSON. It
was separated in three scripts that were called from a Makefile. Some
utility functions were written in Node.js; the node.js module was
called from several scripts.

Our intention was to look for users in a particular location, that is,
to limit them not by minimum number of users but for the location
declared in their profiles. Every run of the program required 10 API
requests which are limited to 20 per hour, so our first modification \cite{Merelo2015}  was to ochange it so that it
used authenticated requests. Finally, we had to rearrange the
whole code so that it counted the number of stars and could be
filtered according to regular expressions, since the country a city or
province is might be ambiguous (you know Toledo, Ohio, but there is
also a Toledo in Spain); Markdown handling was hard-coded into the program 
so it was moved to a template-based solution. The resulting solution
\cite{Merelo2015} kept the same license and included also a few
additional tools for data processing. 

One of the main problem we found in Spain was the different forms of the
province name. Besides the fact that people write it in any of the
official languages (Spanish and, in some cases, national languages
like Basque or Catalan) and, well, sometimes with typos (with or
without tildes), some people do not mention their province when
writing their location. To make a long story short, we had to provide
a configuration file (in JSON) which lists several possible names that
might be used by people in a particular province; for instance, for
Majorca we had to include this: {\tt "location":
  ["Balears","Baleares","Palma de Mallorca"]}. This, of course,
excludes those that simply do not care about listing their location,
but more on this later on.

The script is then run with a city name (if there is no particular
configuration option, {\tt Madrid}, for instance) or a configuration
file name {\tt granada}. This can be launched weekly, or simply at a
particular moment or under request. 

The results for each user list the number of followers, contributions,
the number of stars his/her repositories have received, the longest
and the current 
contribution streak as well as the {\em predominant} language and
avatar. Some of these metrics are shown in the Markdown rankings; for
instance, see the 
\href{https://github.com/JJ/top-github-users-data/blob/master/formatted/top-Madrid.md}{one for Madrid}.

Data is saved to a 
\href{https://github.com/JJ/top-github-users-data/}{different repository} and is aggregated
and processed using R and Perl scripts. All of them are included in
the same repository. As part of our commitment to free/open science,
all graphics and data were published as soon as they were produced in
Twitter from my {\tt @jjmerelo} account. 

\section{Results and analysis}
\label{sec:res}

We reduced the search to the 20 most populated provinces in Spain, for
which appropriate search strings and filters were created\footnote{Population data was obtained from the National Statistics
  Institute \url{http://www.ine.es/}}. All data
was downloaded during January 2015, and is available from the already
mentioned \emph{repo}.  Aggregate data for these 20 provinces is shown in
table \ref{tab:data}.

\begin{figure}[htb]
  \centering
\begin{knitrout}
\definecolor{shadecolor}{rgb}{0.969, 0.969, 0.969}\color{fgcolor}
\includegraphics[width=.49\linewidth]{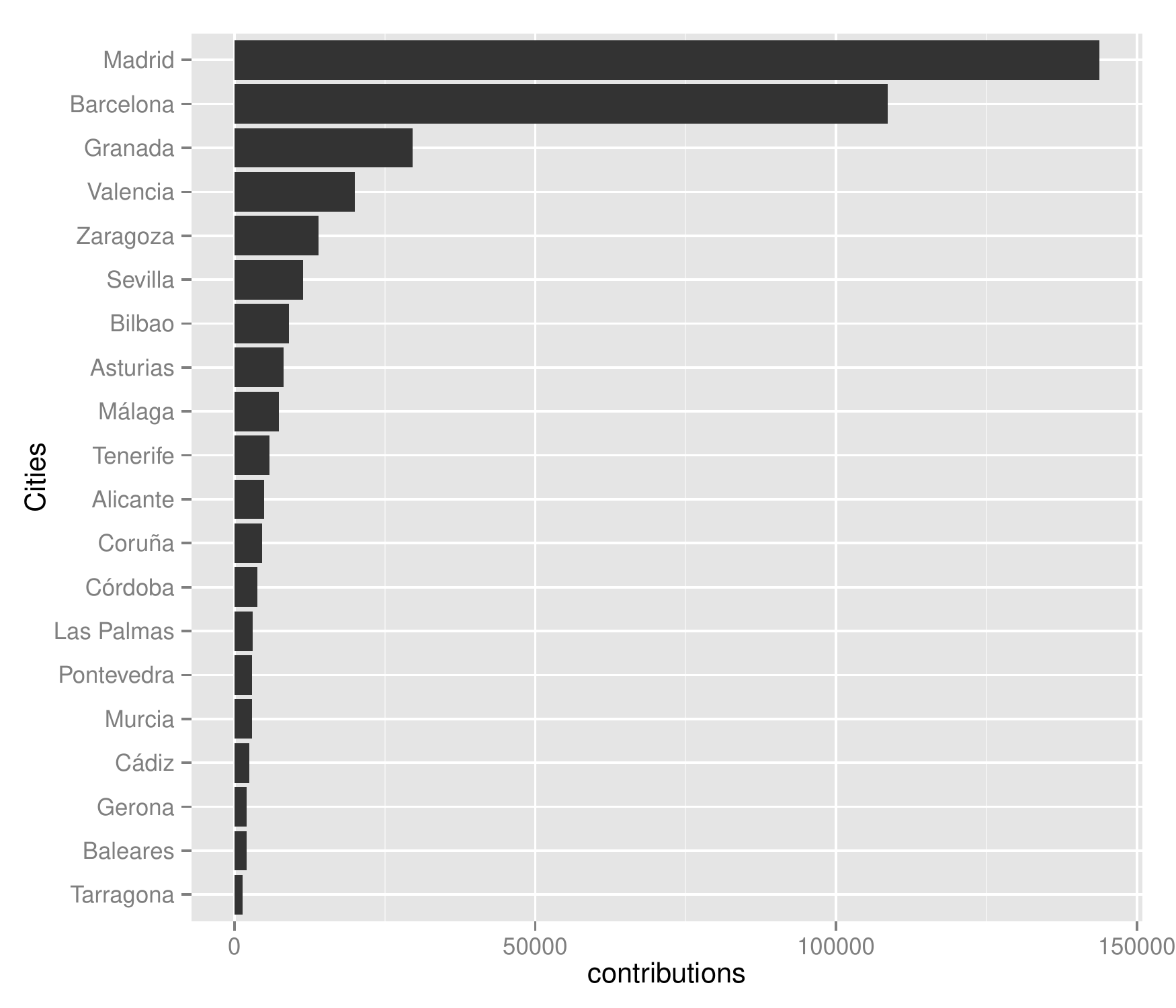} 
\includegraphics[width=.49\linewidth]{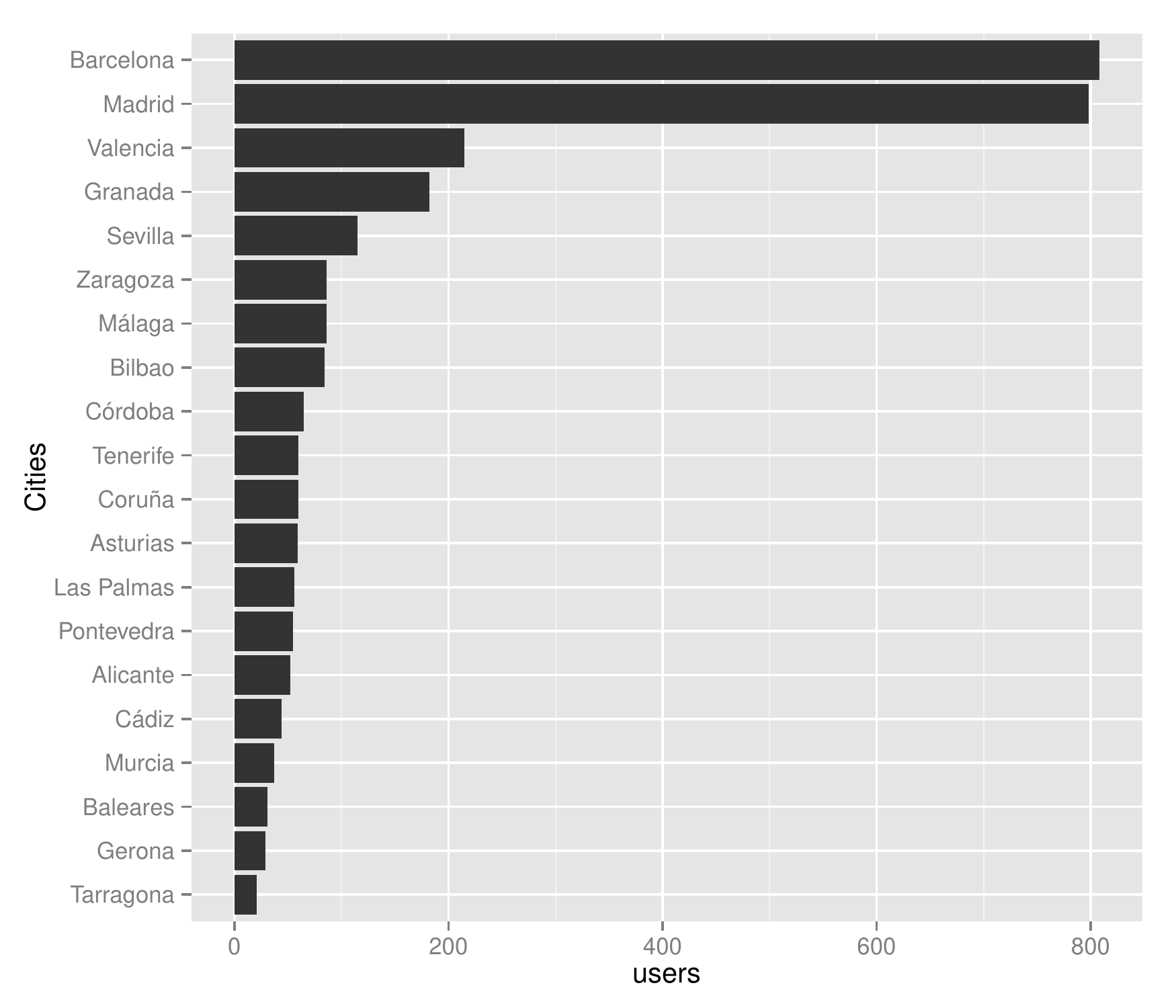} 

\end{knitrout}
\caption{Provinces ranked by number of contributions (left) and users (right)}
\label{fig:user:contrib}
\end{figure}

The range of users shown in the table hovers around the hundreds, with
the one in the biggest provinces (and cities) approaching
1000. However, population and users/contributions are not directly
related. We can already see some differences in figure \ref{fig:user:contrib}, that shows the number of contributions (left) and users (right) in decreasing order. The first two, Madrid and Barcelona, should only be expected, but then Granada (17th in population) and Zaragoza also occupy a place that does not correspond exactly to the population they have; same as Bilbao (actually Vizcaya), but in the opposite direction.  

\begin{figure}[htb]
  \centering
\begin{knitrout}
\definecolor{shadecolor}{rgb}{0.969, 0.969, 0.969}\color{fgcolor}
\includegraphics[width=.49\linewidth]{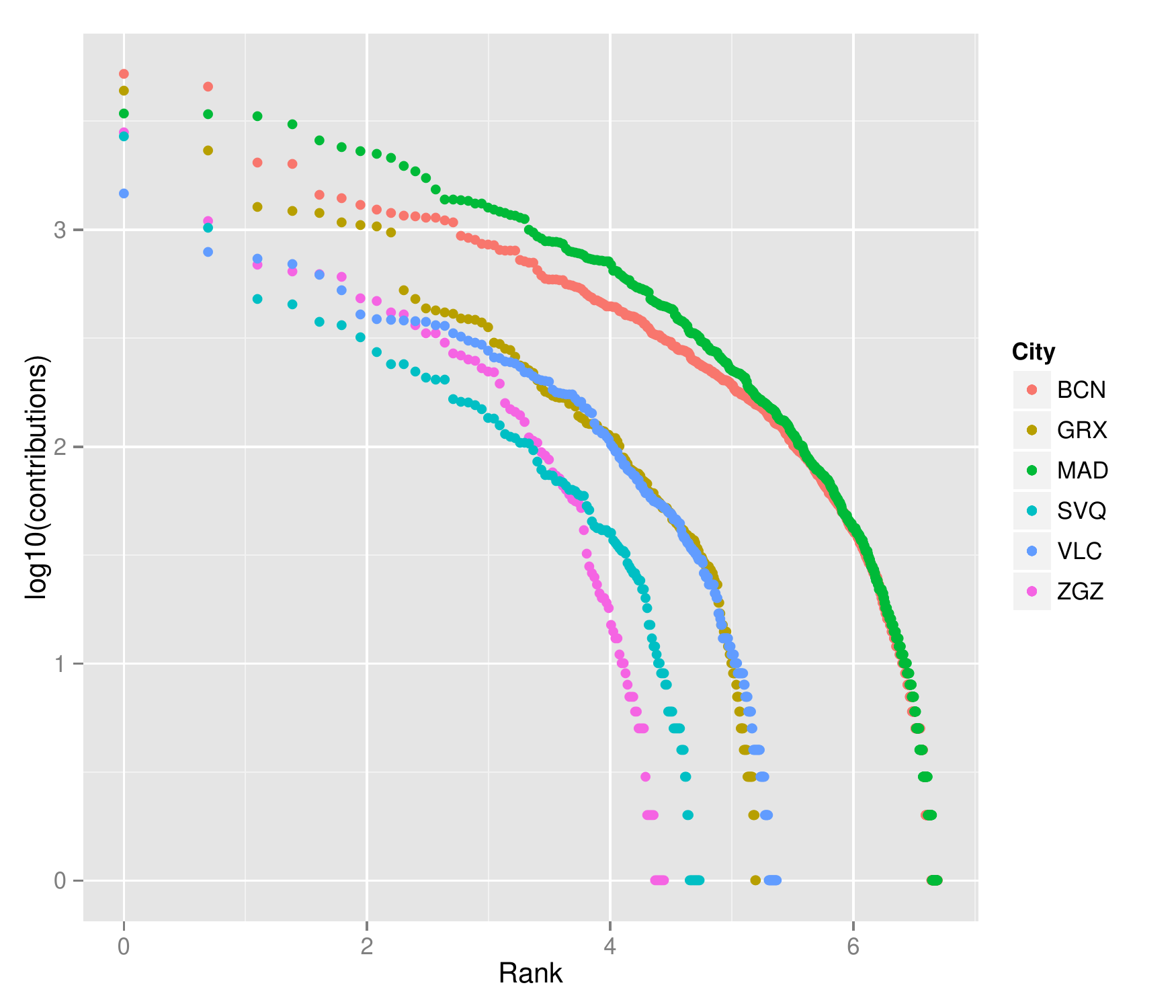} 
\includegraphics[width=.49\linewidth]{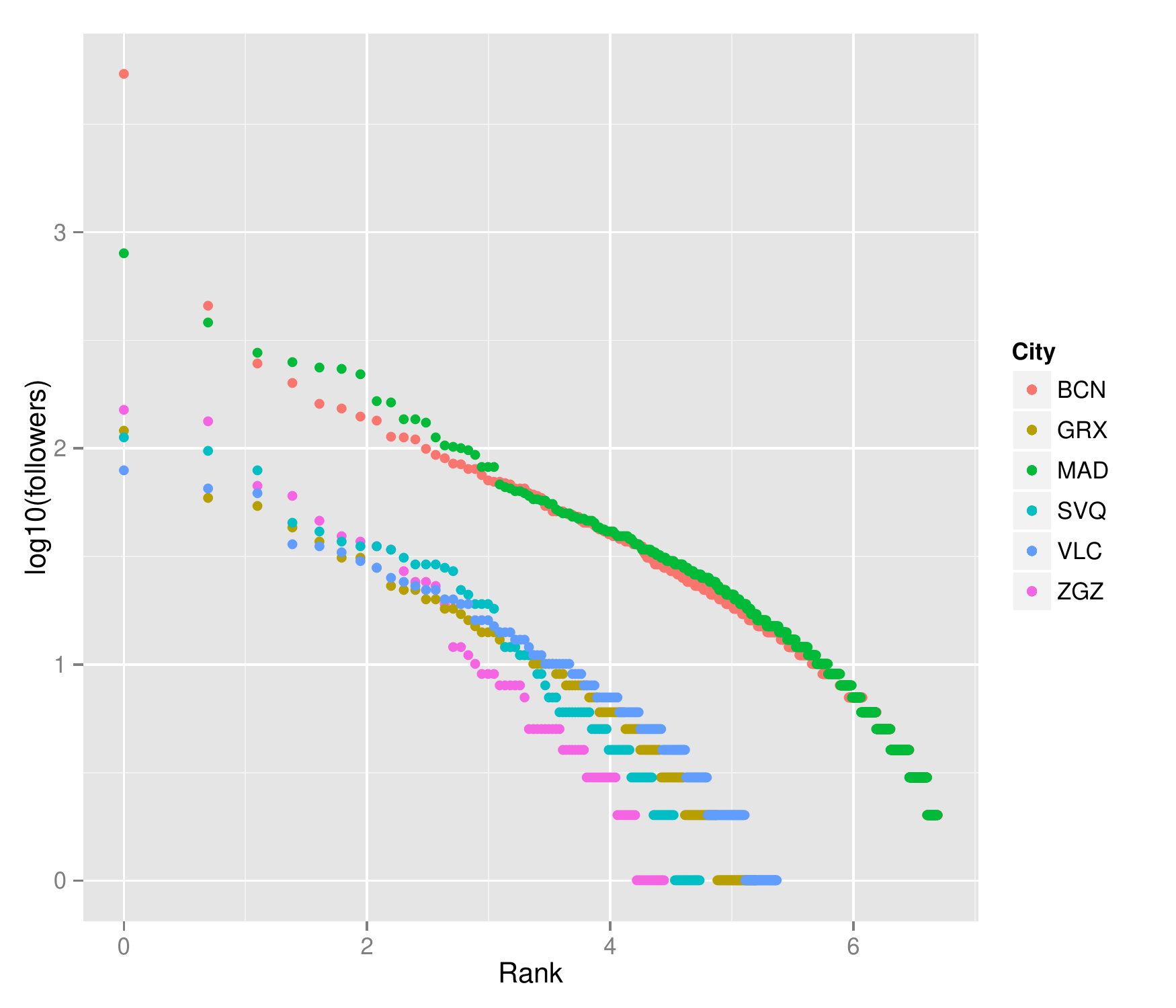} 

\end{knitrout}
\caption{Zipf graph, rank vs. number of contributions (left) or followers (right) for the top 6 provinces in number of users and contributions: Madrid, Barcelona, Valencia, Granada, Sevilla, Zaragoza}
\label{fig:zipf}
\end{figure}

So let us look at the distribution of contributions and users looking for an explanation of the dancing places in the ranking. A users and contributions vs. rank plot is shown in figure \ref{fig:zipf}; it shows different slopes which imply different distribution, but there is a clear indication that a Zipf-like distribution is taking place in all cases. So let us compute the Zipf exponent and objective, which we show in table \ref{tab:zipf}.

\begin{table}[htb]
  \centering

\begin{knitrout}
\definecolor{shadecolor}{rgb}{0.969, 0.969, 0.969}\color{fgcolor}
\begin{tabular}{l|r|r}
\hline
city & exponent & obj\\
\hline
Zaragoza & 1.510043 & 85.85441\\
\hline
Sevilla & 1.342720 & 81.24680\\
\hline
Granada & 1.259239 & 91.25858\\
\hline
Valencia & 1.227201 & 172.91440\\
\hline
Madrid & 1.201127 & 739.38890\\
\hline
Barcelona & 1.157690 & 737.07838\\
\hline
\end{tabular}

\end{knitrout}

\caption{Zipf coefficients for all 6 ``big'' cities, with exponent and objective }
\label{tab:zipf}
\end{table}

\begin{figure}[htb]
  \centering
\begin{knitrout}
\definecolor{shadecolor}{rgb}{0.969, 0.969, 0.969}\color{fgcolor}
\includegraphics[width=\maxwidth]{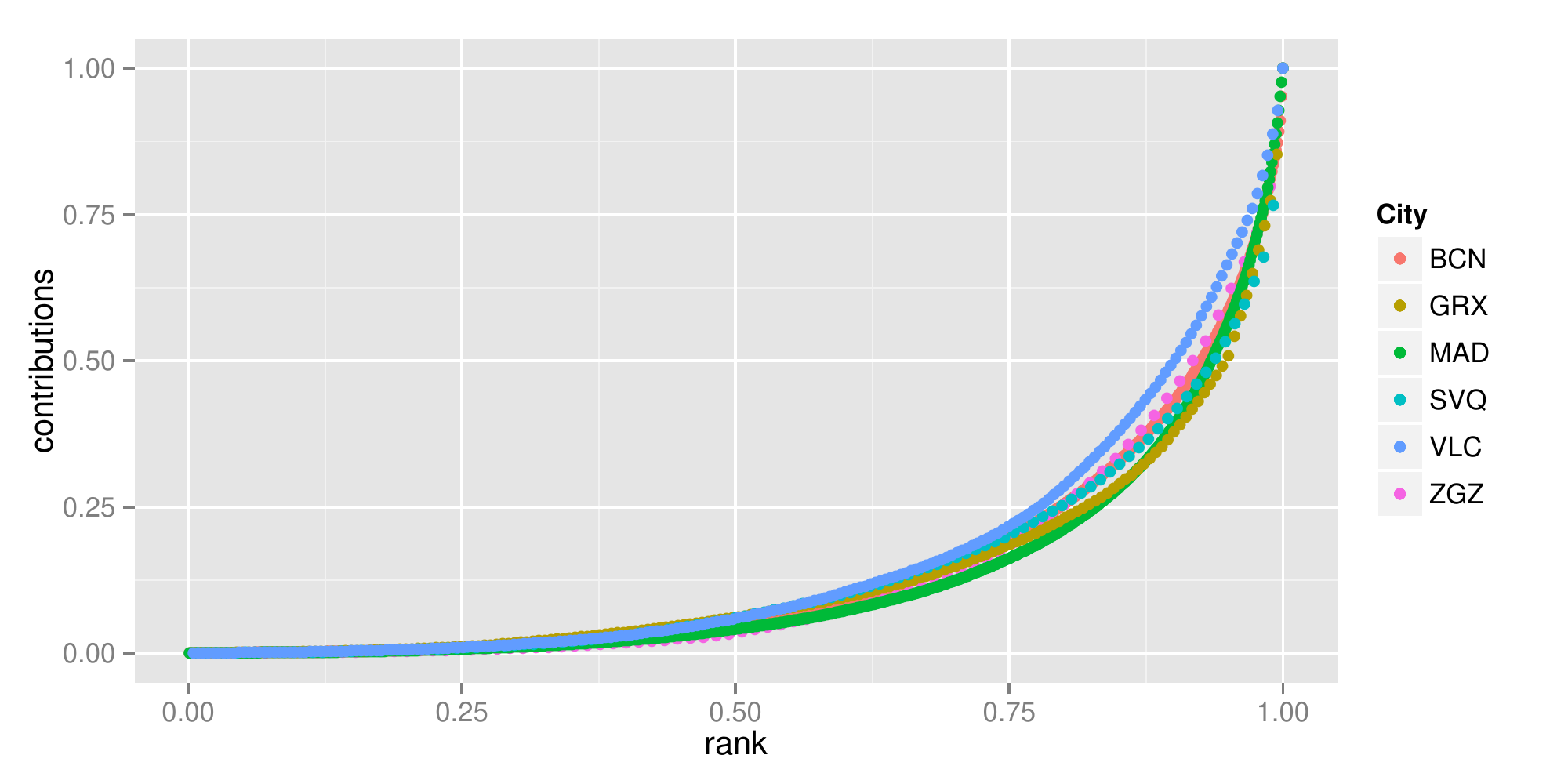} 

\end{knitrout}
\caption{Lorenz graph, that is, accumulated number of contributions vs rank for the top 6 provinces in number of users and contributions: Madrid, Barcelona, Valencia, Granada, Sevilla, Zaragoza}
\label{fig:lorenz}
\end{figure}

This can be interpreted in a different way by plotting the Lorenz
curve, which is the accumulated normalized sum of contributions for
these six cities. This is shown in figure \ref{fig:lorenz}; this
Lorenz curve tends to represent the inequality between those that
contribute more and those that contribute less and is usually
represented by the Gini index, which is shown in table
\ref{tab:gini}. 
\begin{table}[htb]
    \centering

\begin{tabular}{l|r}
\hline
city & gini\\
\hline
Madrid & 0.7491426\\
\hline
Granada & 0.7347302\\
\hline
Zaragoza & 0.7281052\\
\hline
Sevilla & 0.7224632\\
\hline
Barcelona & 0.7206917\\
\hline
Valencia & 0.6839236\\
\hline
\end{tabular}

\caption{Gini coefficients for all 6 ``big'' cities }
\label{tab:gini}
\end{table}

\begin{figure}[htb]
  \centering
\begin{knitrout}
\definecolor{shadecolor}{rgb}{0.969, 0.969, 0.969}\color{fgcolor}
\includegraphics[width=\maxwidth]{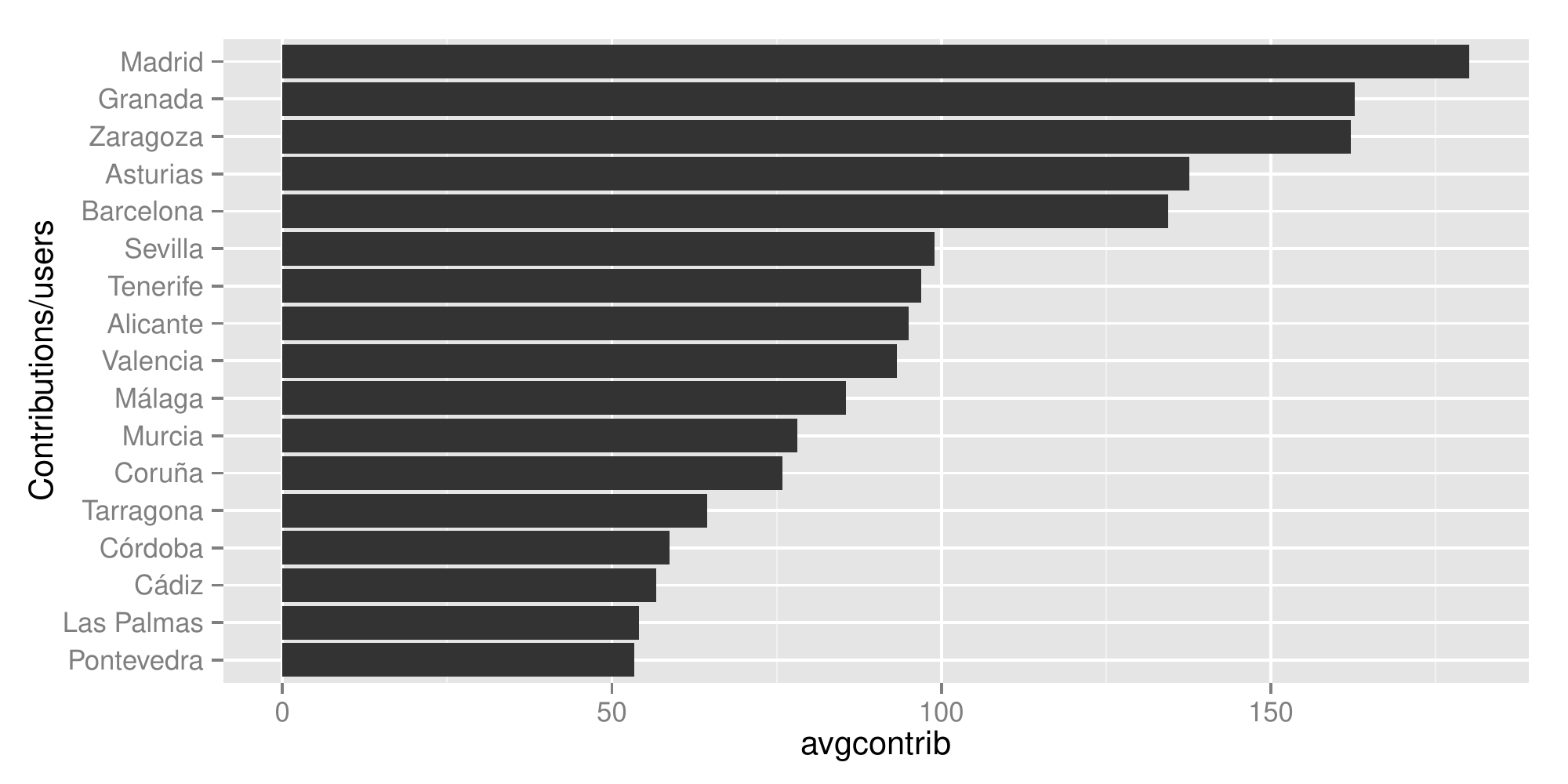} 

\end{knitrout}
\caption{Average number of contributions, sorted from top to bottom,
  for the 20 provinces with the most inhabitants.}
\label{fig:avg}
\end{figure}

The Gini coefficient measures {\em inequality} in the sense of {\em
  share} of, in this case, contributions between those with the most
contributions and those with the least. An index equal to 1 would mean
a single person did all the contributions, while the rest did 0, and
index equal to 0 would mean all users make the same number of
contributions. The table \ref{tab:gini} ranks the cities from least
equal (Madrid) to most egalitarian, Valencia. However, there is no
big range of variation, hovering around 0.70, which is way over the
inequality of the most unequal country in the world, the
Seychelles. However, this is meaningless in absolute terms; in
relative terms, it means roughly that the top contributors contribute
roughly 70\% more than the bottom contributors, and that there is no
big variation among the different cities/provinces. It is quite clear,
however, that the contributions by the top contributor, as well as
those made by the {\em average} one, are quite different from place to
place. So we represent in figure \ref{fig:avg} the average number of
contributions, that is, the number of contributions divided by the
number of users. 

Figure \ref{fig:avg} shows that average contributions have a bigger
range than the Gini coefficient; Valencia is right in the middle,
around 100; in fact, the top contributor ({\tt pakozm} has 1462
contributions and 25\% have more than 100). In Madrid, however, the top
10 have more than 2000 contributions and there are 200 users (25\%
with more than 150), so that accounts for the bigger inequality. But
productivity is highest in Madrid, Zaragoza and, once again, in
Granada, if we consider productivity exclusively the number of
contributions. 

\begin{figure}[htb]
  \centering
\begin{knitrout}
\definecolor{shadecolor}{rgb}{0.969, 0.969, 0.969}\color{fgcolor}
\includegraphics[width=\maxwidth]{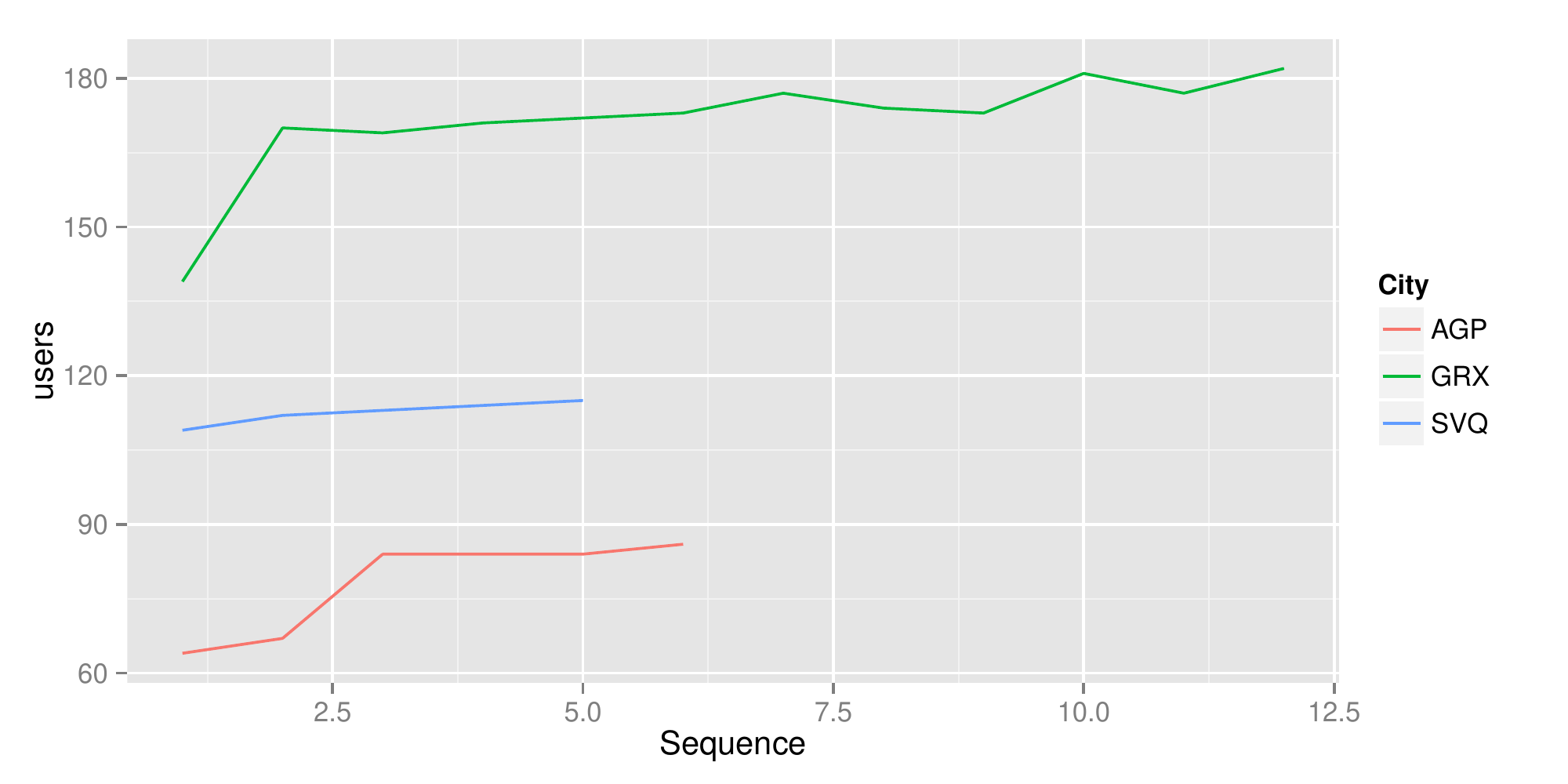} 

\end{knitrout}
\caption{Number of users in each ranking; every point indicates simply
a new ranking and are not uniformly distributed in time.}
\label{fig:impact}
\end{figure}

Finally, it is interesting to find out if the publication of these
rankings had any kind of impact. This is shown in figure
\ref{fig:impact} for the three cities for which we have the most
tests, including Granada.
The first time the census for Granada was published it included around
140 users. A month (approximately) later, there were more than 180, a
28\% increase, more or less the same than for Málaga and more than for
Seville, where a small increase was noted. Take into account that this
is not absolute number of users, but only {\em active} users; in fact,
it might decrease due to some user becoming inactive (no contribution)
in the last year (this happens in some of the cases in Granada). So,
in general, it should be expected to go every which way, depending on
the city. The fact that all cities whose rankings have been published
have {\em increased} the number of active users after diffusion,
mainly in Twitter, might be an indication more users becoming active,
more mentioning their city/province in their profile, or small
competitions taking place locally to scale up the rankings if there is
a chance to do so. All hypothesis are equally valid lacking other
evidence, but we would say that at least some increase will be due to
the fact that the rankings exist. 

\section{Conclusions}

In this paper we have shown what happened when city/province based
rankings were created using GitHub search API and what conclusions can
be extracted from measuring the number of users and contributions made
by these users. 

In general, it is interesting to note that Spain hosts a vibrant,
abundant and diverse community of developers. Looking at the raw
numbers, most of them are based in the big cities, Madrid, Barcelona
and Valencia, but some smaller cities like Zaragoza and Granada also
host a numerous and productive group of developers. The publishing of
the rankings has created a lively discussion in Twitter, and also
allowed the discovery of many developers in many areas. In Granada
GitHub monthly meetings have started, and many interesting projects,
including this paper, have been started. 

There are many things that remain to be done. The first one is to
check the ability of GitHub to act as a social network; we would like
to analyze how developers in a city connect to each other and how
these actual communities change with time and what is their
background, companies, academia or user groups. Other productivity
measures could also be taken, including number of lines; besides, a
differentiation between code and artifacts could be made, in the same
way it was done by Robles et al. in \cite{robles06}.

\section{Acknowledgements}

This paper is part of the open science effort at the university of
Granada. It has been written using knitr, and its source as well as
the data used to create it can be downloaded from
\href{https://github.com/geneura-papers/github-ranking}{the GitHub repository}. It has been supported in part by 
\href{http://geneura.wordpress.com}{GeNeura Team}.

\bibliographystyle{alpha}
\bibliography{geneura,rankings,blogs}

\end{document}